\documentclass[a4paper,twocolumn]{esapub}

\usepackage{graphicx}
\usepackage{times,natbib}

\title{Microquasar-AGN-GRB Connections}
\author{I.F. Mirabel}
\affil{Service d'astrophysique. CEA-Saclay. France.}
\affil{Instituto de Astronom\'{\i}a y F\'{\i}sica del Espacio. CONICET. Argentina.}

\begin{document}

\maketitle


\begin{abstract}
I review the progress made on the physics of  black hole systems in 
the context of the analogy between AGN and microquasars that was proposed 
one decade ago. 
If the emerging empirical correlations between the observational properties 
of stellar and supermassive black holes will become more robust, 
we will use them to determine the mass and spin of black holes, 
independently of theoretical models.  Microquasars are fossils of 
sources of Gamma-ray bursts (GRBs) of long duration, and their kinematics 
provides observational clues on the physics of collapsars.  
If jets in GRBs, microquasars and AGN are due to a unique universal 
magnetohydrodynamic mechanism, synergy between the research on these three 
different classes of cosmic objects will lead to further progress in 
black hole physics and astrophysics.

The potential contributions with INTEGRAL in this field of research are also 
discussed. It is shown that the Galactic Plane Survey (GPS) with INTEGRAL is 
unraveling 
new extragalactic sources of hard x-rays behind the Milky Way, as well as 
new black hole high-mass binaries that were hidden by large 
columns of gas and dust. 
\end{abstract}

\section{Physics in black holes of all mass scales}

The physics in all systems dominated by black holes is essentially the same, 
and  it is governed by the same scaling laws. 
The main differences derive 
from the fact that the scales of length and time of the phenomena 
are proportional to the mass of the black hole. If the lengths, 
masses, accretion rates, and luminosities are expressed in 
units such as the gravitational radius 
(R$_g$ = 2GM/c$^2$), the solar mass, 
and the Eddington luminosity, the same physical laws apply to 
stellar-mass and supermassive black holes \citep{Sams,Rees}. 
For a black hole of mass M the density and mean temperature in the accretion 
flow scale 
with M$^{-1}$ and  M$^{-1/4}$, respectively. For a given
critical accretion rate, the bolometric luminosity and length of relativistic 
jets are proportional to the mass of the black hole. 
The maximum magnetic field at a given radius in a radiation dominated 
accretion disk scales with  M$^{-1/2}$, which implies that 
in the vicinity of stellar-mass black holes the magnetic fields may 
be 10$^4$ times stronger 
than in the vecinity of supermassive black holes \citep{Sams}. 
In this context, it was proposed \citep{Mirabelnat92,Mirabelnat98} that 
supermassive black holes in quasars and stellar-mass black holes in 
x-ray binaries should exhibit analogous phenomena. Based on this 
physical analogy, the word 
``microquasar" \citep{Mirabelnat92} was chosen to designate compact 
x-ray binaries that are sources of relativistic jets (see Figure 1). 

\begin{figure*}
\centering
\includegraphics[width=8cm]{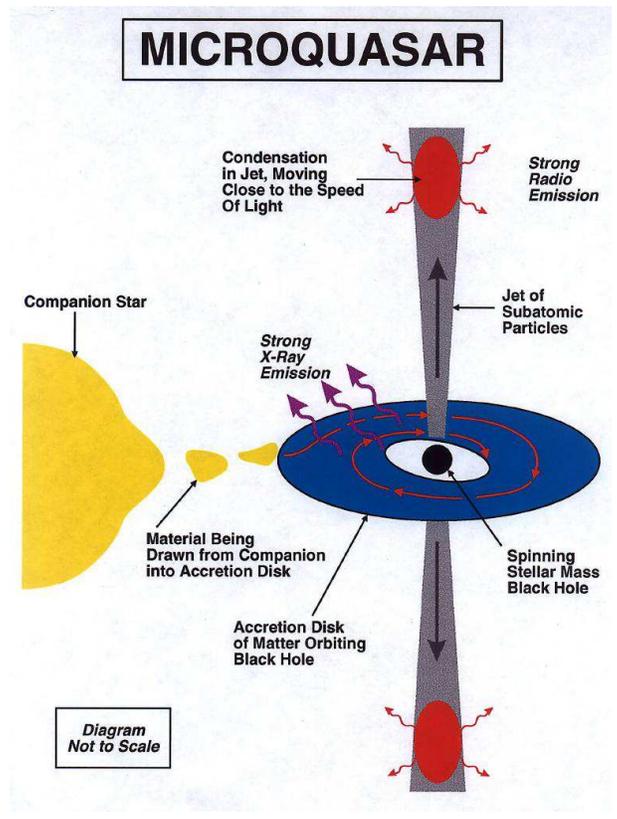}
\caption{This diagram illustrates current ideas of what may be microquasars.  
These are x-ray binary systems that eject plasma at relativistic speeds.}
\label{mylabel1}
\end{figure*}



\section{Superluminal motions in AGN and microquasars}

A galactic superluminal ejection was observed for first 
time in the black hole x-ray binary GRS 1915+105,  
at the time of a sudden drop in the BATSE 20-100 keV flux 
\citep{Mirabelnat94}. 
Since then, relativistic jets with comparable bulk Lorentz factors 
$\Gamma$ = 1/[(1-$\beta$)$^2$)]$^{1/2}$ as 
in quasars have been observed in several other 
x-ray binaries \citep{MirabelARAA,Fender1,Paredes}. At present, it is believed  
that all x-ray accreting black hole binaries are jet sources. 

\begin{figure*}
\centering
\includegraphics[width=12cm]{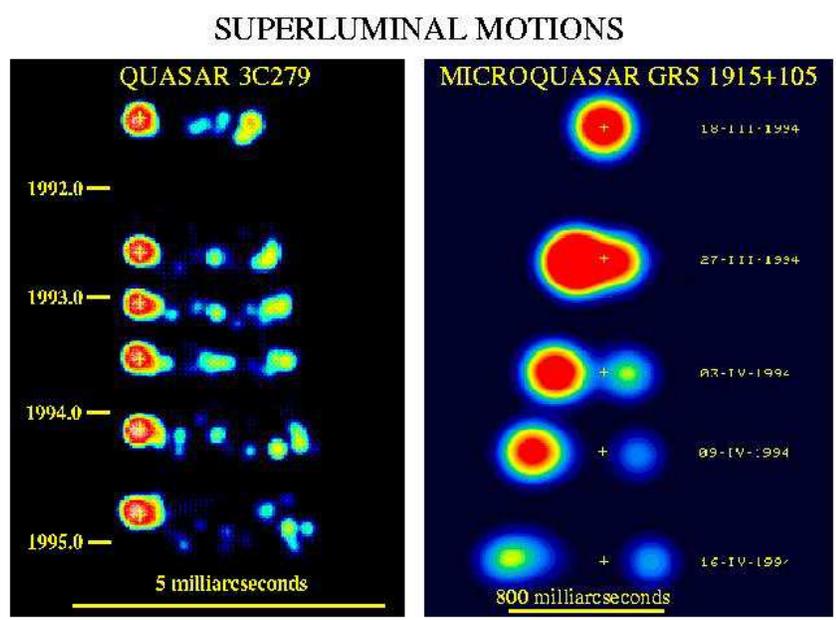}
\caption{Apparent superluminal motions observed  
in the microquasar GRS 1915+105 at 8.6 GHz \citep{Mirabelnat94} and in 
the quasar 3C 279 at 22 GHz.}
\label{mylabel2}
\end{figure*}

Galactic microquasar jets usually move in the plane of the sky 
$\sim$10$^3$ 
times faster than quasar jets and can be followed 
more easily than the later  (see Figure 2). Because of their proximity, in microquasars two-sided jets 
can be observed, which together with the distance provides the necessary data to solve the system of 
equations, gaining insight on the actual speed of the ejecta. 
On the other hand, 
in AGN located at 
$\leq$100 Mpc, the jets can be imaged with resolutions of a few times 
the gravitational radius of the supermassive black hole, as was done for 
M 87 \citep{Biretta}. This is not 
presently possible in microquasars, since such a precision in terms of 
the gravitational radius of a stellar-mass black hole would 
require resolutions a few hundreds of kilometers. Then, in terms of the 
gravitational radius
in AGN we may learn better  
how the jets are collimated close to the central engine. In summary, 
some aspects of the relativistic jet phenomena associated to accreting 
black holes are better observed in 
AGN, whereas others can be better studied in microquasars. 
Therefore, to gain 
insight into the physics of relativistic jets in the universe, synergy 
between knowledge of galactic and extragalactic black hole is needed.

\section{Accretion-jet connection in microquasars and quasars}

Microquasars have allowed to gain insight into the connection 
between accretion disk instabilities and the formation of jets.  
In $\sim$1 hour of simultaneous multiwavelength observations of GRS 1915+105 during the frequently observed 30-40 min x-ray oscillations in this 
source, the connection between sudden drops of 
the x-ray flux from the accretion disk and the onset 
of jets were observed in several ocassions \citep{MirabelAA98,Eikenberry} 
(see Figure 3).

\begin{figure*}
\centering
\includegraphics[width=14cm]{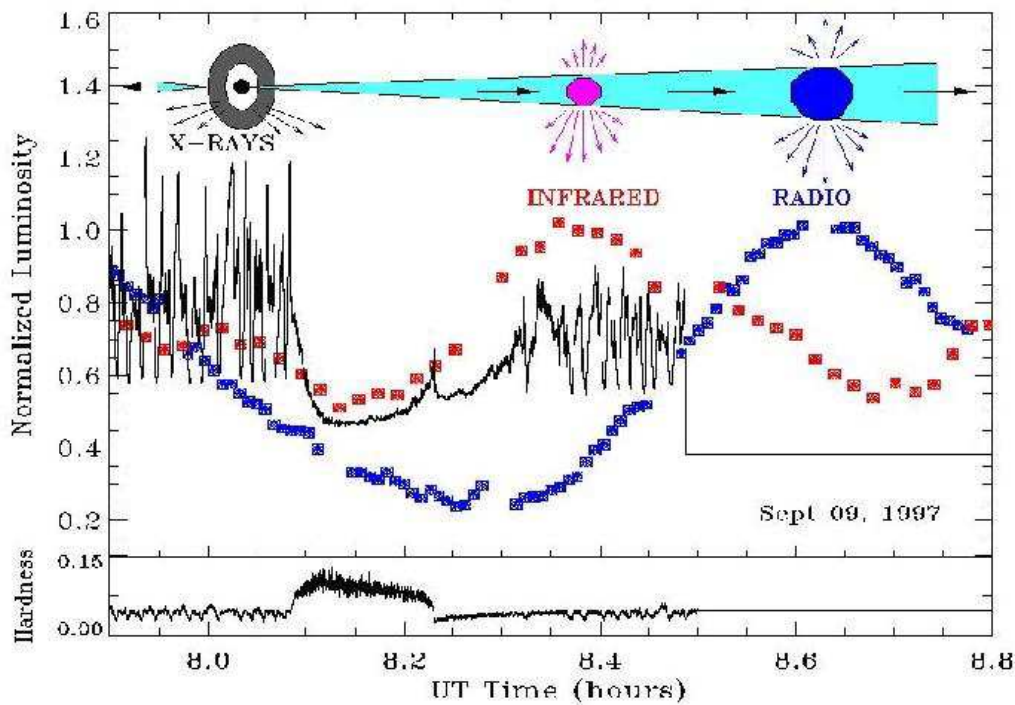}
\caption{Direct evidence for the disk-jet connection in the black hole x-ray 
binary GRS 1915+105 \citep{MirabelAA98}. When the hot inner accretion disk disappeared, its x-ray brightness abruptly diminished. The ensuing x-ray recovery documented the inner disk's replenishment, while the rising infrared and radio emission showed plasma being ejected in a jet-forming episode. 
The sequence of events shows 
that material indeed was transfered 
from the disk to the jets. Similar transitions have been observed in the 
quasar 3C 120 \citep{Marscher}, but in time scales of years, rather than minutes.}
\label{mylabel3}
\end{figure*}

From these observations we have learned the following:

a) the jets appear after the drop of the x-ray flux, 

b) the jets are produced during the replenishment of the inner 
accretion disk, 
 
c) the jet injection is not instantaneous. It can last up to 
$\sim$10 min,
 
d) the time delay between the jet flares at wavelengths of 2$\mu$m, 
2cm, 3.6cm, 6cm, and 21cm are consistent with the model of 
adiabatically expanding clouds that had been proposed to account 
for relativistic jets in AGN \citep{vanderLaan},

e) synchrotron emission is observed up to infrared wavelengths and 
probably up to x-rays. This would imply the presence in the jets of electrons with TeV energies,

f) VLBA images during this type of x-ray oscillations \citep{Dhawan} showed that 
the ejecta consist on compact collimated jets with lengths of $\sim$100 AU.

g) there is a time delay of $\sim$5 
min between the large drop of the x-ray flux from the accretion 
disk and the onset of the jets. These $\sim$5 minutes of silence suggest 
that the compact object in GRS 1915+105 has a space-time border, rather than a material border, namely, a horizon as expected in 
relativistic black holes. However, the absence of evidence of 
a material surface in these observations could have alternative explanations.

After the observation of this accretion disk-jet connection in a microquasar, 
an analogous connection was 
observed in the quasar 3C 120 \citep{Marscher}, but in scales of 
years rather than minutes. This time scale ratio is 
comparable to the mass ratio between the supermassive black 
hole in 3C 120 and the stellar black hole in GRS 1915+105, as expected 
in the context of the black hole analogy. 

\section{X-ray jets in AGN and microquasars}

X-ray emission has been observed with Chandra and XMM-Newton in the 
radio jets, lobes and hot spots of quasars and radio galaxies. 
X-ray photons 
can be produced by inverse Compton scattering from the environment of 
the central engine up to distances of $\sim$100 kpc \citep{Wilson}. 
Synchrotron x-ray radiation has been 
detected in some sources, most notably in the jet of M 87. 

Steady, large-scale radio jets are associated to  
x-ray persistent microquasars \citep{MirabelARAA}.  
Extended x-ray emission associated to the radio emission was observed in 
the galactic source SS433/W50 up to distances of $\sim$30 pc from 
the central engine \citep{Dubner} (see Figure 4). The x-rays extend  
in the same direction as the sub-arcsec, precessing jets. 
The jets in SS433 are hadronic, move with a velocity of 0.26c, 
and have a kinetic power of $\sim$10$^{39-40}$ erg s$^{-1}$. 
Since no hot spots have 
been detected in the shock regions 30 pc away from the central 
engine, SS433/W50 cannot be considered a scale-down analog of a 
FR II radio galaxy. Most microquasars with extended jet emission 
have morphologies analogous to FR I's rather than FR II's radiogalaxies. 

\begin{figure*}
\centering
\includegraphics[width=14cm]{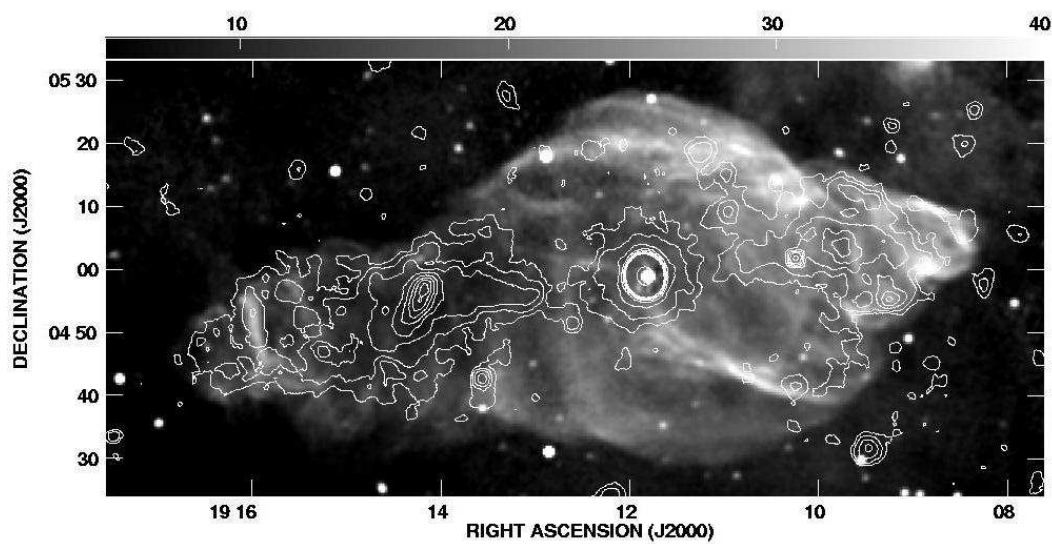}
\caption{VLA radio image at 1.5 GHz (grey scale) superimposed on the ROSAT 
x-ray contours of SS433/W50 \citep{Dubner}. The radio counterpart of the microquasar is 
the bright unresolved source at the center of the image. The lateral E-W extension of the nebula over $\sim$1$^{\circ}$ ($\sim$50 pc) is caused by 
the injection of relativistic jets from SS 433.}
\label{mylabel4}
\end{figure*}

The recent discovery of radio \citep{Hjellming} and x-ray \citep{Corbel} 
moving jets in  
microquasars rise the possibility of studying the formation 
of radio and x-ray lobes in real time. These observations show that jets 
may transport energy in a ``dark" way, namely, in a way that is radiatively  inefficient, until shocks are produced. Synchrotron x-ray emission 
from shocks at large distances from the central 
engines imply that microquasars are potential sources of cosmic rays and 
electrons with up to TeV energies.

\section{Blazars and microblazars}

The bulk Lorentz factors $\Gamma$ in microquasars and quasars 
have similar values, 
and it had been proposed \citep{MirabelARAA} that microquasars with 
jet axis that form angles $\leq$10$^{\circ}$ with the line of sight 
should appear as ``microblazars", showing analogous phenomena
to blazars. Due to relativistic beaming, 
in microblazars the brightness of outbursts is enhanced by factors of 
8 $\times$ $\Gamma$$^3$ and the interval of time of the phenomena 
is reduced 
by factors of 1/2 $\times$ $\Gamma$$^{-2}$. Then, microblazars 
should appear as intense sources 
of high energy photons with very fast variations of flux, which makes 
them difficult to find and to follow. Due to this difficulty and to 
the relatively low statistical probability of small angles between the jet axis 
and the line of sight \citep{MirabelARAA}, it is not surprising that most 
of the 
microquasars studied so far exhibit large angles ($\geq$30$^{\circ}$) between the jet axis and the line of sight.

It has been proposed that microblazars may be more frequently  
found in High Mass X-ray Binaries (HMXBs) \citep{Paredes2,Romero}. 
In such binaries, gamma-rays can be produced by inverse Compton 
of the jet particles with the UV photons radiated by the massive donnor 
star. In fact, the three microquasars so far proposed as counterparts 
of variable EGRET unidentified sources are HMXBs with similar 
properties \citep{Combi}.

\section{Extragalactic microquasars and super-Eddington x-ray sources}

GRS 1915+105 and SS 433 have been invoked  \cite{King} as the Milky Way counterparts 
of the two most numerous classes of super-Eddington x-ray sources 
(ULXs) in external galaxies. GRS 1915+105 is a long lasting 
transient outburst x-ray binary with an evolved donnor of $\sim$1 M$_{\odot}$, 
whereas SS 433 is a persistent black hole HMXB \citep{Hillwig}. 
SS 433 type of ULX's are preponderantly 
found in starburst galaxies like the Antennae, whereas luminous x-ray 
sources of low mass as GRS 1915+105 may also be found in galaxies with 
a low rate of star formation. 

Most of the ULX's would be stellar-mass black hole microquasars with the following possible 
properties:
 
1) HMXBs that host massive stellar black holes (M $\geq$ 40 M$_{\odot}$) with  isotropic 
radiation \citep{Pakull}.

2) HMXBs and LMXBs that host stellar black holes (M $\sim$ 10 M$_{\odot}$) 
with anisotropic radiation \citep{King}. 

3) A few ($\leq$1\%) may be microquasars with relativistic boosted radiation. 
These should be very bright, highly time-variable, and have a hard x-ray/$\gamma$-ray photon 
spectrum. \citep{MirabelARAA}.

Although less numerous, it is not exculded that some ULX's could be 
accreting black holes of intermediate-mass (100-1000 M$_{\odot}$).

\section{X-ray/Radio correlations in low power black holes of all masses}

Several teams of researchers are exploring interesting 
x-ray/radio correlations. 

Microquasars in the low-hard state exhibit radio/x-ray 
correlations \citep{Gallo}. In the low-hard state the power output 
of quiescent black holes  
is jet-dominated and when the system moves to a high soft state 
the radio jets are quenched. The same seems to take place 
in AGN \citep{Maccarone}.     
A scheme to unify low-power accreting black holes has 
been proposed \citep{Falcke}, where the black holes in Sgr A$^*$, 
LINERs, FR I, and BL Lac would be analogous to microquasar black holes 
in the low-hard state. 

Following studies \citep{Heinz} of correlations between radio and 
bolometric luminosities, a fundamental plane of black hole activity in terms of the black hole 
mass and x-ray and radio core luminosities is 
proposed \citep{Merloni}. This correlation holds 
for radiativelly inefficient accretion, not for bright thin 
synchrotron emitting states. 

At present, these empirical correlations have large scatters. 
However, if they became more robust, the mass of black holes 
could be inferred from the x-ray and radio fluxes, 
independently of theoretical models.

\section{Time variations of flux and the masses of black holes}

Time variations of flux may be correlated  
with the mass of the black hole. 

1) The duration of the x-ray flares observed in stellar-mass black 
holes and in Sgr A* seem to be proportional to the mass of the black 
holes. In Cygnus X-1 and other x-ray black hole binaries, flares with 
durations of 1-10 ms are observed \citep{Gierlinski}. On the other 
hand in Sgr A$^*$, x-ray flares lasting 400-10,000 sec have been 
observed with Chandra \citep{Baganoff} and XMM-Newton \citep{Goldwurm}. As 
expected, the time ratios of the power variabilities are comparable to the black hole mass ratios. 

2) For a given black hole spin, the  maximum frequencies of quasi 
periodic oscillations (QPOs) of flux are expected to be proportional 
to the mass of the black hole. In 4 microquasars, 3:2 twin 
peak x-ray QPOs 
of maximum frequency in the range of 100-500 Hz have been observed, from which angular momenta a = J/(GM/c$^2$) = 0.6-0.9 
have been derived \citep{Abramowicz}.  On the other hand, 17 min infrared 
QPOs have been reported in Sgr A$^*$, from which it has 
been inferred an angular momentum a = 0.52 \citep{Genzel}.  
As expected, these QPOs appear to scale with the mass of the black 
hole. If the 17 min QPO in Sgr A$^*$ is confirmed as a 
component of a twin peak fix QPO of maximum frequency, 
this correlation could be used to 
derive black hole masses, and in particular, those of the super-Eddington 
x-ray sources in external galaxies \citep{Abramowicz}. 

3) Some properties of the aperiodic variability (noise) in AGN and x-ray 
binaries seem to be correlated with the mass of the compact objects. 
The break time scale in the power spectra density of black holes 
seems to scale linearly with the mass of the black hole \citep{Uttley}. 
The broad band break time in the Sey 1 NGC 3516 scales linearly with that 
of Cyg X-1 in the 
low-hard state \citep{McHardy}. If this type of correlation is confirmed it 
could also be used to estimate the mass of black holes 
in extragalactic super-Eddington x-ray sources.

\section{Relativistic iron lines in stellar and supermassive black holes}

AGN frequently exhibit broad iron K$\alpha$ lines skewed to low 
energies \citep{Tanaka}. The shape of these lines is consistent 
with emission from the surface of an accretion disk extending from about 
6 to more than 40 gravitational radii. Occasionaly the red wing of the 
line extends below 4 keV and the current explanation is that the disk 
extends within 6 gravitational radii implying that the black hole is rapidly spinning. Now it is widely believed that this spectral feature 
is a probe of the immediate environment of black holes \citep{Fabian}. 
Until recently, only smeared edges with little evidence for line emission 
had been observed in Galactic black hole binaries. But after 
Chandra, XMM-Newton and Beppo-SAX, similar emission iron lines 
to those in AGN were found, even in the ASCA archive \citep{Miller}. 

Besides emission lines skewed to low energies, analogous spectra to AGN-like warm 
absorbers  are observed in some x-ray binaries \citep{Miller2}. 
The absorption is variable and it is believed to be produced in a 
dense local disk wind rather than in the ISM. A finding possibly 
related to these absorption lines is the discovery with INTEGRAL of black 
hole binaries with strong x-ray 
absorption, much larger than that derived from  
optical and infrared observations, which also implies that the absorption 
is local rather than in the ISM. 

The iron line in stellar black hole binaries can be used 
to investigate: 

1) the physical models of the K$_{\alpha}$ line. Because of the short 
dynamical time scales, the shape of the line can be correlated 
with the x-ray state of the accretion disk, and corona-disk interactions.
 
2) the dense plasma outflows from accreting black holes. 
The study of warm absorber lines  similar to those seen in Seyferts 
may be important to estimate the mass outflows in x-ray binaries. 

3) the spin of the accreting black hole. This is important to test 
models where the jets are powered by the spin of the black hole.

At present, the main constrain to derive the slope of the iron lines is due to the uncertainties on the shape of the continuum at energies 
$\geq$8 keV.

\section{Microquasars and ULX's as fossils of gamma-ray burst sources}

GRBs of long duration may be caused by the formation of a black holes 
in collapsars \citep{MacFadyen} or highly magnetized neutron stars 
\citep{Vietri}. 
It is estimated that half of the compact objects are produced in primordial 
binaries and that after their formation a significant fraction ($\sim$20\%) 
remain in binary systems \citep{Belczynski},
 leaving microquasars as remnants.

It is believed that GRBs take place in close massive binaries because: 

1) the core must be spun up by spin-orbit interaction in order to provide 
enough power to the jet that will drill  the collapsing star 
all the way from the core up to the external layers 
\cite{Izzard,Podsiadlowski}.

2) GRBs seem to be associated to SNe Ic. This is the class of SNe that 
do not show H and He lines, implying that before the explosion 
the progenitor of those GRBs had lost the H and He layers. These 
layers are more easily lost if the progenitor was part 
of a massive binary that underwent a common envelope phase 
\cite{Izzard,Podsiadlowski}. 
Furthermore, SNe Ic exhibit 4-7 \% polarization, which are an indication 
of asymmetric explosions probably caused by collimated jets. 

\section{Microquasar kinematics and the core-collapse}

It is believed that stellar black holes can be formed in two different
ways: Either the massive star collapses directly into a black hole
without an energetic supernova explosion, or an explosion occurs in a
protoneutron star, but the energy is too low to completely unbind the
stellar envelope, and a large fraction of it falls back onto the
short-lived neutron star, leading to the delayed formation of a black
hole \citep{Fryer}.
 If the collapsar takes place in a binary that  
remains bound, and the core collapse produced an energetic 
supernova, it will impart the center of mass of the system with a 
runaway velocity, no matter the explosion being symmetric or 
asymmetric. Therefore, the kinematics of microquasars can be used 
to constrain theoretical models on the explosion of massive stars that  
form black holes. 

Using this method it has been shown that the x-ray binary Cygnux X-1 
was formed in situ and did not receive an energetic trigger from 
a nearby supernova \citep{MirabelSci}.
 If the progenitor of the 
black hole and its parent association Cygnus OB3 are coeval, 
the progenitor mass was greater than 40 M$_{\odot}$, and during the 
collapse to form the 10 M$_{\odot}$ black hole of Cygnus X-1, 
the upper limit for the mass that could have been suddenly ejected 
is $\sim$1 M$_{\odot}$, much less than the mass ejected in a typical supernova. 
Theoretical models suggest that larger mass remnants are associated to 
subluminous supernovae \cite{Balberg,Fryer}. 

Furthermore, the kinematics of GRS 1915+105 derived from VLBA proper motion of the 
compact jet for the last 5 years \citep{Dhawan04}
 show that the  
14$\pm$4 M$_{\odot}$ black hole \citep{Greiner} in this x-ray binary probably was 
formed promptly, as the black hole in Cygnus X-1. Of course, these observations 
do not exclude the possibility that high mass stellar black holes could also 
be formed with strong natal kicks, and runaway as unbound solitary black holes, 
which would be difficult to detect.

On the other hand, the kinematics of GRO J1655-40 has been a confirmation 
\citep{MirabelAA2}
of the indirect evidence for an energetic supernova explosion in the 
formation of this low-mass black hole, inferred earlier 
from the chemical composition of the donnor star \citep{Israelian}. 
More recently, the observation of the runaway x-ray binaries LS 5039 \citep{Ribo}
and LSI +61$^{\circ}$303 \citep{Mirabel04}
confirmed that neutron stars may be formed with strong natal kicks.

\begin{figure*}
\centering
\includegraphics[width=14cm]{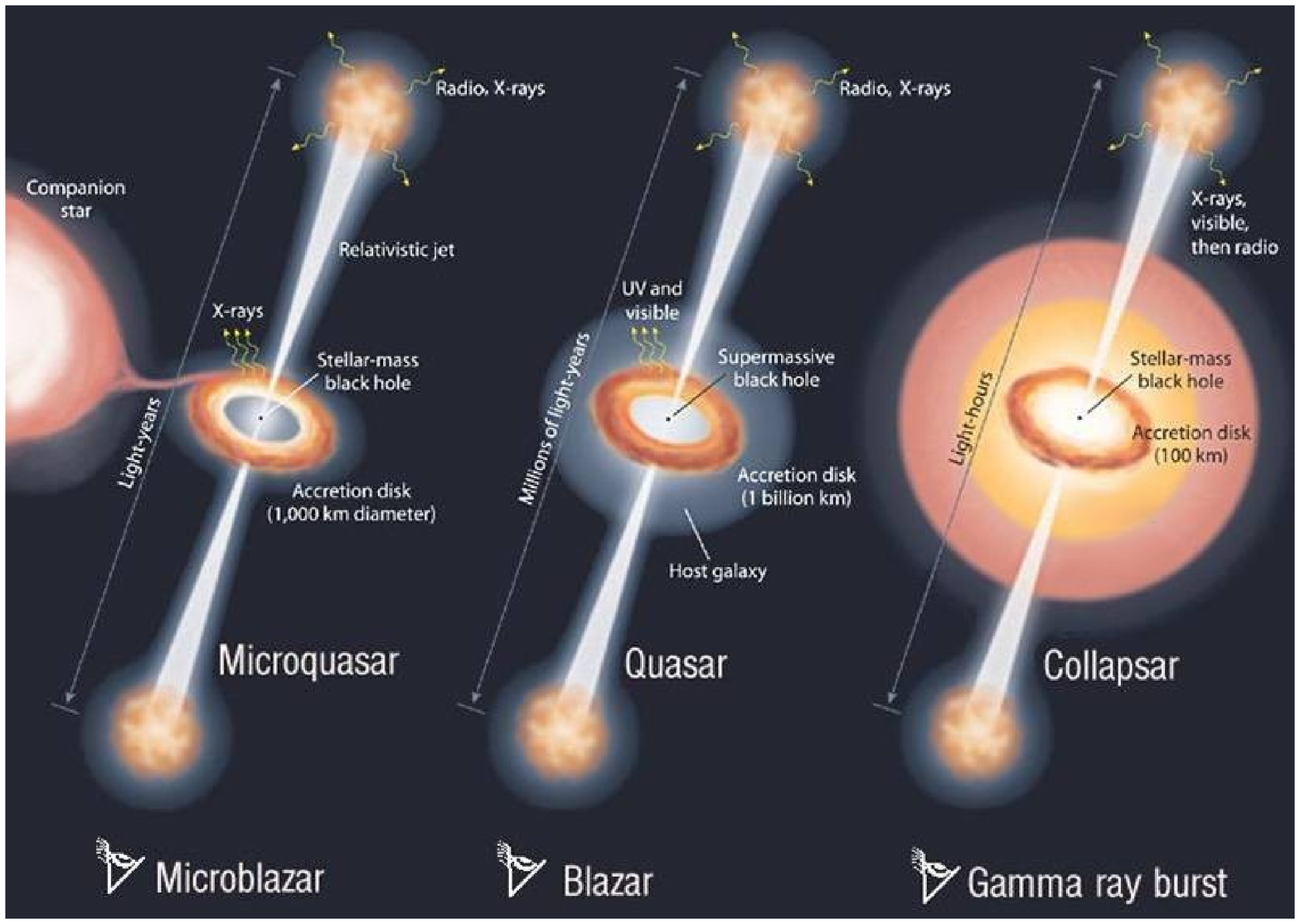}
\caption{A unique universal mechanism may be responsible for three types of 
astronomical objects: microquasars (left); quasars (center); and 
collapsars (right), the massive, suddenly collapsing stars believed to cause some gamma-ray bursts. Each contains a black hole (probably spinning), an accretion disk (which transfers material to the black holes), and relativistic jets (which emerge from a region just outside the black holes, carrung away angular momentum). Microquasars and quasars can eject matter many times, while collapsars form jets but once. When the jet is aligned with an observer's line of sight these objects appear as microblazars, blazars, and gamma-ray bursters, respectively. The components of each panel are not drawn to scale; scale bars denote jet lengths. (Sky \& Telescope, May 2002, 32)}
\label{mylabel5}
\end{figure*}

\begin{figure*}
\centering
\includegraphics[width=8cm]{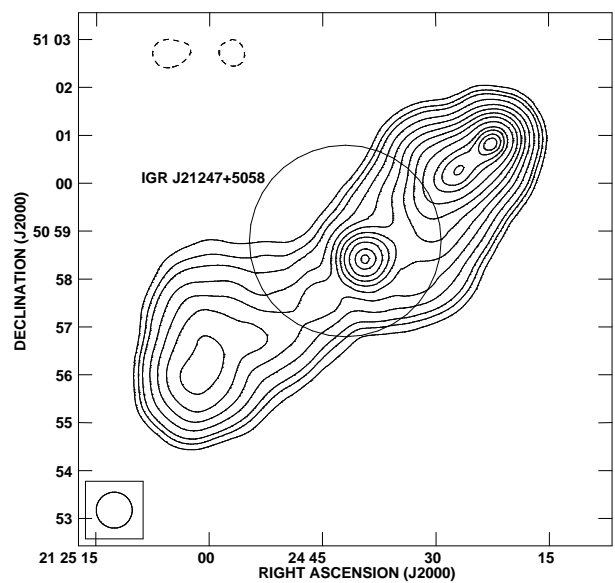}
\caption{NVSS image at 1.4 GHz of the radiogalaxy 4C 50.55 detected in the Galactic 
Plane Survey of INTEGRAL as IGR J21247+5058. The core of the radio source is within 
the 90\% uncertainty error circle of the hard x-ray source \cite{Ribo04}.}
\label{mylabel5}
\end{figure*}

\section{Progenitors of neutron stars}

It is commonly believed that massive stars with masses above and below 
a fix mass limit leave black holes and neutron stars, respectively. 
However, the destiny of a massive star depends on several other factors besides on the mass, 
such as the metallicity and whether it is solitary or in a binary. 

More massive stars evolve to the supernova stage faster, and a lower limit 
for the mass of the progenitor of the compact object can be 
determined assuming that the progenitor was coeval with the stars 
of the parent cluster. 

It has been found that the neutron star x-ray binary LSI +61$^{\circ}$303 was 
shot out from its birth place in the cluster 
of massive stars IC 1805 and that the progenitor of the compact object may have 
had a mass $\geq$ 56 M$_{\odot}$ \citep{Mirabel04}.
However, this cluster is in a complex region 
and a lower mass progenitor cannot be ruled out since sequential star formation 
could have taken place in that region. 

The location of 
magnetars (soft gamma-ray repeaters) in clusters of massive stars also 
sugest that neutron stars may have very massive stellar progenitors. 
Two of the four known magnetars (SGR 1806-20 and SGR 1900+14) are inside 
dust-enshrouded clusters of massive stars \citep{MirabelSGR}.


\section{Relativistic jets in AGN, microquasars and GRB sources}

It was suggested that irrespective of their 
mass there may be a unique universal mechanism for 
relativistic jets in accreting black holes \citep{MirabelST} (see Figure 5). 
Although in AGN, microquasars and GRB sources there are different 
physical conditions, it was proposed  
that all jets are produced by an unique electromagnetic mechanism, in which 
charged plasma is accelerated by electric fields that are generated 
by a rotating magnetic field \citep{Meier}. However, the most popular GRB 
jet models at this time are baryon dominated, 
and the factors that control 
the jet power, collimation and speed, remain unknown.

\section{New black holes in the INTEGRAL GPS}
 
Besides the targeted observations of known black hole binaries that are reported in 
other 
contributions of these proceedings, the GPS survey is revealing new hard x-ray sources 
with spectral properties that are characteristic of accreting black holes. The observed hard x-ray 
properties of stellar mass and supermassive black holes are very similar and for obscured 
sources it is usually needed to determine the  
redshift of the counterpart to know whether the hard x-ray source is a compact object of 
stellar mass in a binary or an AGN. 

In the GPS with INTEGRAL several hard x-ray sources were discovered 
hidden behind large columns of interstellar gas and dust \citep{Bird}. 
Multiwavelength follow up of these sources is showing that some are dust enshrouded 
high mass black hole binaries, such as IGR J16318-4848 \citep{Chaty} and 
AX J1639.0-4642 = IGR J16393-4643 \citep{Combi}. It is interesting that 
IGR J16393-4643 is, as the microquasars LS 5059 and LSI +61$^{\circ}$303, 
inside the error circles of EGRET sources of high-energy gamma-ray emission, 
and that these three microquasars have similar properties \citep{Combi}.

The number of detections of AGN in the GPS is smaller 
than expected on the basis of the sensitivity of INTEGRAL, and some of the newly 
discovered hard x-ray sources may be extragalactic \cite{Bassani}. In fact, after the 
discovery with GRANAT 
of a Seyfert 1 galaxy at a redshift of 0.021 behind the Galactic Center \cite{Marti}, 
INTEGRAL is detecting new hard x-ray AGN. Figure 6 shows that IGR J21247+5058 
is the hard x-ray counterpart of the radiogalaxy 4C 50.55 \citep{Ribo04}. 
The source IGR J18027-1455 is probably also an AGN, because the photometric 
properties of the infrared and optical counterpart are not consistent with 
those of a stellar object \citep{Combi2}. The multiwavelength approach to the 
hard x-ray sources revealed by INTEGRAL is in progress.

\section*{Acknowledgements}
\noindent For this review I have greatly benefited from work done  
in collaboration with   
L.F. Rodr\'\i guez, I. Rodrigues,  M. Rib\'o, S. Chaty, J. Mart\'{\i}, V. Dhawan,  
R. Mignani, J. Combi and L. Pellizza.

\end{document}